\def\c12{$^{12}$C}
\def\c12l{$^{12}_\Lambda$C}

\def\kmno16{$^{16}{\rm O}(K^-, n)$}
\def\kmNc12{$^{12}{\rm C}(K^-, N)$}
\def\kmnc12{$^{12}{\rm C}(K^-, n)$}
\def\kmpc12{$^{12}{\rm C}(K^-, p)$}
\documentclass[letter]{ptptex}

\usepackage{xcolor}            
\usepackage{graphicx}
\usepackage{subfigure}         
\notypesetlogo                       

\title
{Search for a bound state of kaon and pion}  

\author{
T. \textsc{Kishimoto}$^{1,2}$%
\  F. \textsc{Khanam}$^1$, \ T. \textsc{Hayakawa}$^1$, \ S. \textsc{Ajimura}$^2$, \
T. \textsc{Itabashi}$^1$, \ K. \textsc{Matsuoka}$^1$, \ S. \textsc{Minami}$^1$, \ 
Y. \textsc{Mitoma}$^1$, \ A. \textsc{Sakaguchi}$^1$, 
\ Y. \textsc{Shimizu}$^4$, \ K. \textsc{Terai}$^1$, \ T. \textsc{Sato}$^1$, 
\ H. \textsc{Noumi}$^{2,3}$, \ M. \textsc{Sekimoto}$^3$, \ H. \textsc{Takahashi}$^3$, \ 
T. \textsc{Fukuda}$^4$, \ W. \textsc{Imoto}$^4$ \ and Y. \textsc{Mizoi}$^4$ %
}
 
\inst{
$^1$Department Physics, Osaka University, 
Toyonaka, Osaka 560-0043, Japan\\
$^2$Research Center for Nuclear Physics (RCNP), Osaka University, Ibaraki, 
Osaka, 560-0047 Japan\\
$^3$High Energy Accelerator Research Organization (KEK), Tsukuba, 
305-0801, Japan\\
$^4$Osaka Electro-Communication University, Neyagawa, 572-8530, Japan
}

\abst{
We have searched for a bound state of kaon and pion denoted by $X$.  The $X$ 
was conjectured to explain the so-called $\Theta^+$ resonance as a bound state 
of kaon, pion and nucleon.  This model explains almost all properties of the 
$\Theta^+$, however, the model works only 
if the $K \pi$ interaction is strongly attractive.  It is so strong that 
it could make a bound state $X$.  Here we report a result of the search 
for the $X$ by using the $K^+ + N \rightarrow X^+ + N$ reaction at P$_K\sim$ 
1.2 GeV/c.  The $X^+ \rightarrow K^+ \gamma \gamma$ decay produces $K^+$ in 
momentum region where other processes cannot fill.  We observed signature of 
the $X^+$ with statistical significance of 2 $\sigma$.  Production 
cross section of $X$ with respect to that of $\pi^0$ is 1$\pm$0.5\% if we take 
it as an evidence and 1.5\% if we set an upper limit.  
}

\begin{document}

\maketitle

In this paper we present a result of a search for a hypothetical particle as a 
bound state of kaon and pion which was called $X$ \cite{TK_TS}.  The $X$ was 
conjectured to explain the $\Theta^+$ as a bound state of kaon, pion and 
nucleon.  The $\Theta^+$ was first observed by the LEPS collaboration at 
SPring8 as a narrow resonance in the $K^+ n$ invariant mass distribution 
\cite{nakano}.  
It should consist of at least 5 quarks thus natural explanation is a 
penta-quark state.  Many following experiments showed positive evidence on 
the $\Theta^+$ which indicate that its mass is around 1540 MeV 
and width could be less than 1 MeV \cite{arndt-ps}.  On the other hand, 
many negative results also appeared.   Recently the LEPS group disclosed 
a new result which further strengthen the positive evidence \cite{PRC_nakano}.  
They employed the deuterium target where the Fermi momentum correction is 
much smaller than the carbon target in the first experiment.  It is 
not our intention to review vast amount of studies on the $\Theta^+$ which 
can be found elsewhere \cite{PRC_nakano}.  

There are already number of experiments which gave positive and negative 
results, therefore a new single experiment can hardly be decisive on the 
existence of the $\Theta^+$.  Situation may be changed if one presents evidence 
that could consistently explain why certain experiments give positive result 
and others do not.  There is a tendency that positive results are mostly from 
the low energy experiments and negative ones are from the high energy 
experiments.  In particular, no positive results appeared in the collider 
experiments.  One may restate this observation that reactions with higher 
momentum transfer have smaller cross sections.  This observation contradicts 
to the model that the $\Theta^+$ is a penta-quark state.  Since its size is 
similar to that of hadrons (baryons and mesons), one expects that 
the production cross section of the $\Theta^+$ has momentum transfer 
dependence similar to hadrons.  

There are attempts to explain the $\Theta^+$ as a bound state of $K \pi N$ 
\cite{TK_TS, 7quark, Estrada}.  This bound state model gives the spin-parity 
of ${\frac{1}{2}}^+$ same as the original theoretical prediction based on chiral 
soliton model \cite{Diakonov}.  The chiral soliton model has huge pion 
component which is similar to the bound state model.  Mass of three particles 
(1570 MeV) and binding energy of 10 MeV/particle gives the observed 
$\Theta^+$ mass.  In particular, the narrow width of 1 MeV or less can be 
naturally explained in this model \cite{TK_TS}.  The radius of a bound state, 
which depends on its binding energy, is generally much larger than that of 
hadrons.  The cross section is thus expected to be small for reactions with 
high momentum transfer.  This model could give consistent description of both 
positive and negative experimental results.  Recent DIANA experiment gave 
positive result and obtained 0.39 MeV for the width \cite{diana}.  
Such narrow width indicates small production cross section and makes 
experimental studies difficult.  Theoretically, the three boday bound 
state model is almost unique possibility to reproduce such narrow width. 

However, currently known two body interactions of $\pi N$, $K N$ and $K \pi$ 
are not attractive enough to realize the bound state \cite{TK_TS, 7quark, Estrada}.   
The $\pi N$ \cite{arndt-pin} and $K N$ \cite{dover-kn, hashimoto, arndt-kn} 
interactions are based on scattering experiments thus they are rather firm.  
On the other hand, $K \pi$ interaction is based only on production experiments 
since both kaon and pion are unstable particles.  Kishimoto and Sato conjectured 
that the $K \pi$ interaction might be attractive more than that has been known 
\cite{estabrooks} to realize $\Theta^+$ as the bound state and it 
could be strong even to make their bound state $X$ \cite{TK_TS}.  
The lowest order decay mode is $X\rightarrow K \gamma \gamma$ since it is a 
$0^+$ to $0^-$ transition \cite{TK_TS}.  This could be a reason why it has not 
been observed even if it exists.  A straightforward experiment to identify 
the $X$ is to observe invariant mass of a kaon and two $\gamma$'s. 
The experiment requires dedicated detectors in particular $\gamma$ detector.  
Since the $X$ is currently speculative to deserve such a sophisticated 
experiment, we designed an experiment to observe kaon momentum.  

We employed the $N(K^+,X^+)N$ reaction for the production of the 
$X$\cite{TK_TS}.  Its momentum transfer could be even lower than 100 MeV/c 
for the $X$ ejected at forward angles for $P_K\sim$1 GeV/c \cite{TK_TS}.  
We can expect appreciable cross section for the reaction with such small 
momentum transfer.  A $K^+$ momentum spectrum can give 
signature since the $K^+$ from the decay of the $X^+$ could have higher 
momentum than that of $K^+$ induced by $\pi^0$ production reaction.

The experiment (PS-E548) was carried out at the K2 beam line of the 12 GeV 
proton synchrotron of KEK (KEK-PS) \cite{e548}.  The primarily purpose of 
the E548 experiment is a study of kaonic nuclei by the $(K^-, N)$ reaction 
where 1 GeV/c $K^-$ beam was employed.  Since details of the experiment 
have been described in ref \cite{TK_PTP}, we present here what is relevant 
to the present experiment.   The $X$ is quadruplet due to its strangeness 
($s=1$ and $s=-1$) and isospin 1/2.  We could choose either $K^-$ or $K^+$ beam 
to produce $X^-$ or $X^+$, respectively.  We employed 1.2 GeV/c $K^+$ beam 
for the present search due to reasons in the following.  The beam intensity 
of $K^+$ is higher than that of 
$K^-$ and higher beam momentum gives higher beam intensity.  The $K^+$ beam 
gives less background than that of $K^-$ which accompanies hyperon production.  

The beam intensity was typically 4$\times$10$^4$/spill for the primary 
proton beam intensity of $2 \times 10^{12}$/spill.  A spill consisted of 
1.7 second of continuous beam in every 4 
second.  The momentum bite of the incident $K^+$ beam was 17 MeV/c ($\sigma$).  
The momentum of the outgoing $K^+$ was measured by the KURAMA spectrometer 
whose momentum resolution was $\sim$6.4 MeV/c.  
Polyethylene (CH$_2$) was used as a target which had dimensions of 
$10 \times 4.5 \times 20.5$ cm$^3$.  Overall momentum resolution is found to be 
24 MeV/c which includes the energy loss struggling of $K^+$ in the target. 
The target was sandwiched by six 1 cm thick plastic scintillator hodoscopes 
with 5 cm granularity in the $z$ (beam) direction.   $\gamma$ rays from the $X$ 
decay were measured by a decay counter surrounding the target.  The decay 
counter consists of two sets of 25 NaI arrays.  Each NaI has dimensions of 
$6.5 \times 6.5 \times 30$ cm$^3$.  The array was set at 15 cm above and below 
the target.  In front of the NaI detectors, 1 cm thick 
plastic scintillators were placed to identify charged particles.  Additional 12 
NaI counters were placed to cover forward angles which gave supplemental 
information.  
Scattered $K^+$'s at forward angles hardly distinguished from beam through events.  
We set the trigger condition that the NaI arrays have energy deposit more than 
4 MeV.  This reduces trigger rate tolerable to data acquisition system.  
Energy calibration of NaI counters were made by punch through 
muon beam and cosmic ray muons.  In order to obtain $\gamma$ ray energy, 
we first select a NaI which has the maximum energy and summed energy of 
adjacent NaI's. 
The NaI arrays cover $\sim$30\% of solid angle which is insufficient to 
construct invariant mass of $X$ or $\pi^0$.  However, it is enough to tag 
events that accompany $\gamma$ rays for the present experiment.  

\begin{figure}[ht]
 \begin{center}
  \includegraphics[height=90mm]{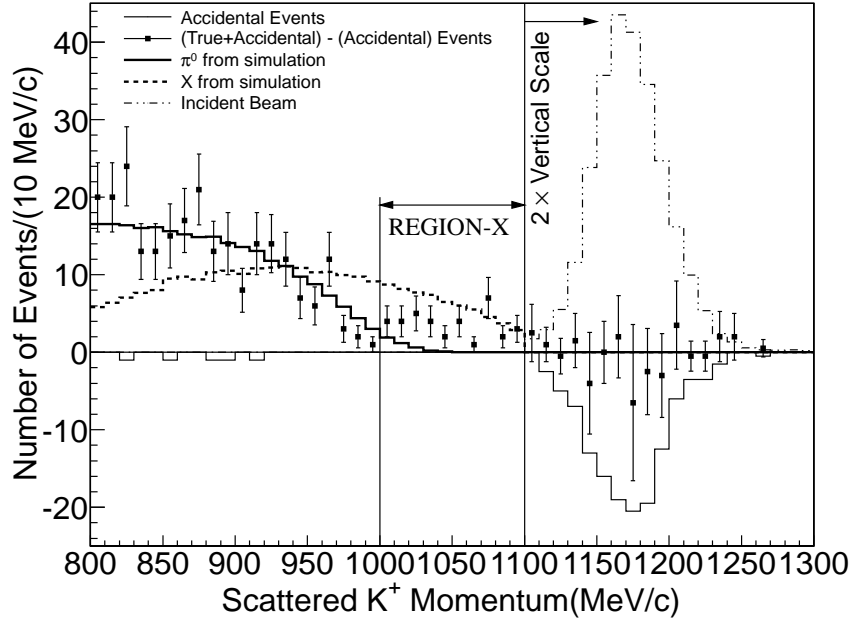}
 \end{center}
 \caption{
Momentum spectrum of the $(K^+, K^+)$ reaction on the CH$_2$ target gated 
by $\gamma$ rays is shown.  Filled boxes with error bars represent 
``true = (true+accidental) - accidental'' events, solid line represents 
``accidental'' 
events which is plotted in the negative Y-direction.  The ``true'' 
events together with the simulated events of $\pi^0$ (thick 
solid line) and $X$(dashed line) are plotted, where both the simulated 
histograms are normalized to the ``true'' data points.  The factor 2 is 
multiplied to the vertical scale above 1100 MeV/c. The incident 
beam from the beam through events are also shown in the figure
which is normalized arbitrary to fit within the vertical scale.
The highest number of events is around 2200 at the peak in the incident
beam histogram.
}
\label{fig:x_sim_data}
\end{figure}

Figure \ref{fig:x_sim_data} shows the $K^+$ momentum spectrum.  Here scattering 
angles of $K^+$'s are from 2 to 12 degrees and accompanying $\gamma$ ray energy 
is more than 10 MeV where no nuclear $\gamma$ rays are expected effectively.  
We took 5 ns for the ``true+accidental'' coincidence and adjacent 2.5 ns for 
both sides for ``accidental'' coincidence.  In order to obtain  ``true'' 
events, ``accidental'' events were subtracted from the ``true+accidental'' 
events.   
The $K^+$ momentum spectrum of the beam through events has a peak at 
1.17 GeV/c.  The large fluctuation of events at around 1.17 GeV/c is 
due to accidental coincidence of vast amount of beam through events.  
We see a bump that starts at around 1000 MeV/c toward low momentum 
region.  This corresponds to $\pi^0$ production.  We have events in a 
1000$\sim$1100 MeV/c region which cannot be due to either accidental 
coincidence nor $\pi^0$ production.  Figure \ref{fig:x_sim_data} shows 
the simulated spectra of $\pi^0$ events and $X$ particle events plotted 
together with the true events. 
The events in 1000-1100 MeV/c region (REGION-X here after) indicates 
production of the $X$ if we can confirm that no background processes 
can fill the REGION-X.  We considered the following processes.  

Firstly, effect of the momentum resolution and momentum spread of the 
beam are evaluated.  It might leak the $K^+$ from the $\pi^0$ production 
into the REGION-X.  The effect is evaluated by the momentum spread of 
beam through events measured simultaneously.  It represents overall 
effect of momentum resolution of the system, momentum spread of the beam, 
energy struggling of the beam in the target and all other relevant effects.  
We include the measured beam momentum distribution in our simulation and 
then calculated $K^+$ momentum spectrum from the 
$p + K^+ \rightarrow p + K^+ \pi^0$ reaction.  In the simulation we assumed 
that the momentum distribution of three particles in the final state 
follows the three body phase space volume.  We included the angular 
coverage of the NaI detector in the simulation to reproduce the effect 
of $\pi^0$ momentum direction on the spectrum although its effect is 
small.  The simulation reproduces well the spectrum below P$_K=$1 GeV/c 
and negligible events are seen above that as shown in figure 
\ref{fig:x_sim_data}. 
The simulated histogram of $\pi^0$ production in figure \ref{fig:x_sim_data} 
is normalized to the ``true'' histogram. The 
integrated number of events of the ``true'' histogram are taken as 
the normalization parameter to which each simulated histogram is 
normalized.

\begin{figure}[ht]
 \begin{center}
  \includegraphics[height=90mm]{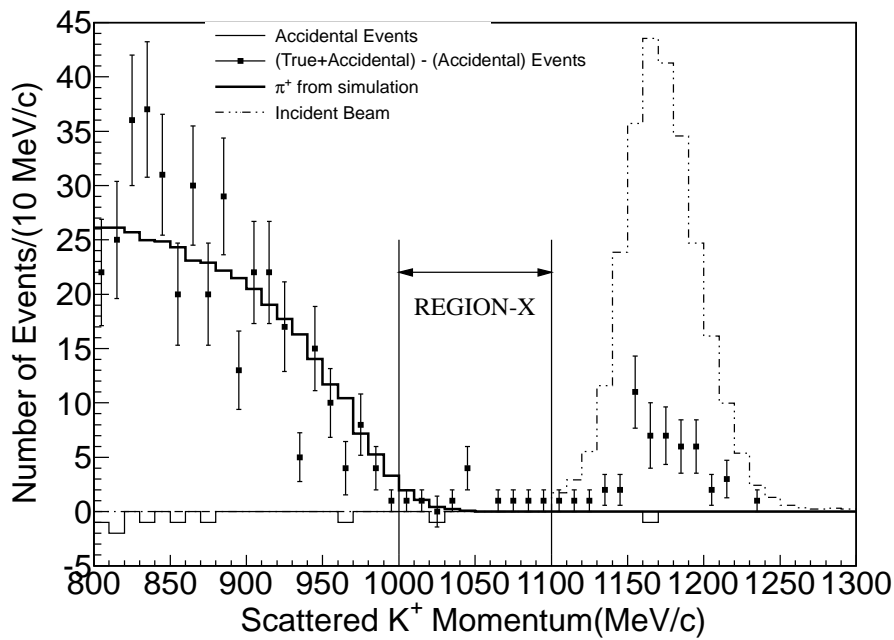}
 \end{center}
 \caption{
Momentum spectrum of the $(K^+, K^+)$ reaction on the CH$_2$ target gated 
by the charged particles is shown.  Data points were shown by filled boxes with 
error bars.  They are results of ``true = (true+accidental) - accidental''.  
The spectrum in the negative Y-direction represents ``accidental'' events.  
The ``true'' events and simulated events of charged pions (thick line) 
are plotted together, where the simulation is normalized to the 
``true'' data points.  The incident beam from the beam through 
events are also shown in the figure.
 }
\label{fig:cp_sim_data}
\end{figure}

Secondly, the reaction takes place not only on a free proton but also on 
nucleons bound in $^{12}$C.  We included Fermi motion of a target nucleon 
to calculate $K^+$ momentum spectrum.  We used 250 MeV/c for the Fermi 
momentum.  Here energy momentum of a nucleon is taken as 
$(E_N, {\mbox{\boldmath $p$}}_F)$ where $ {\mbox{\boldmath $p$}}_F$ is 
three momentum vector of Fermi motion and $E_N$ is an energy of bound 
nucleon given by, 
\begin{equation}
        E_N=m_N-BE-\frac{P_F^2 }{ 2 M_R}, 
\end{equation}
where $BE$ stands for binding energy of a nucleon in a specific state and 
$\frac{P_F^2 }{ 2 M_R}$ is a recoil energy of a residual nucleus.  
It makes the slope at the threshold region ($\sim$1000 MeV/c) gentler 
than that on a nucleon target a little bit.  It slightly reduces events 
in REGION-X but no distinguishable change is seen 
thus cannot explain the events in the REGION-X.

Thirdly, we considered $K^+$ inelastic scattering which may accompany $\gamma$
rays from nuclear excited states.   The REGION-X corresponds excitation 
energy of $60\sim150$ MeV in $^{12}$C.  It is unlikely that inelastic 
scattering excite such region since quasifree scatteing  at forward angles 
can excite up to a few ten's MeV region.  Even if the REGION-X is excited, one 
expects almost no accompanying nuclear $\gamma$ rays over 10 MeV.  
Contrary, emission of particles (protons and neutrons) are expected.  We used 
the NaI detector to tag charged particles by taking coincidence with plastic 
scintilllators in front of the NaI.  Figure \ref{fig:cp_sim_data} shows a 
spectrum gated by charged particles.  The spectrum shows extra events in 
beam through region with a few ten's MeV/c lower momentum region which 
corresponds excitation of $^{12}$C by a few ten's MeV.   We see negligible 
events in the REGION-X.  Since the $X$ has no charged particle decay mode, 
the fact that REGION-X has almost no events in figure 2 and 
some events in figure 1 is consistent with the existence of $X$. 
The spectrum below 1000 MeV/c corresponds reaction with $\pi^+$ production.  
The simulated histogram normalized to the ``true'' data points is shown 
in figure 2.  Good agreement of the simulation with the data indicates that 
we reasonably include relevant processes in the simulation.

\begin{figure}[ht]
 \begin{center}
\subfigure[Mass spectra of the scattered particles.]{
  \label{fig:signal_mass}
  \includegraphics[height=50mm]{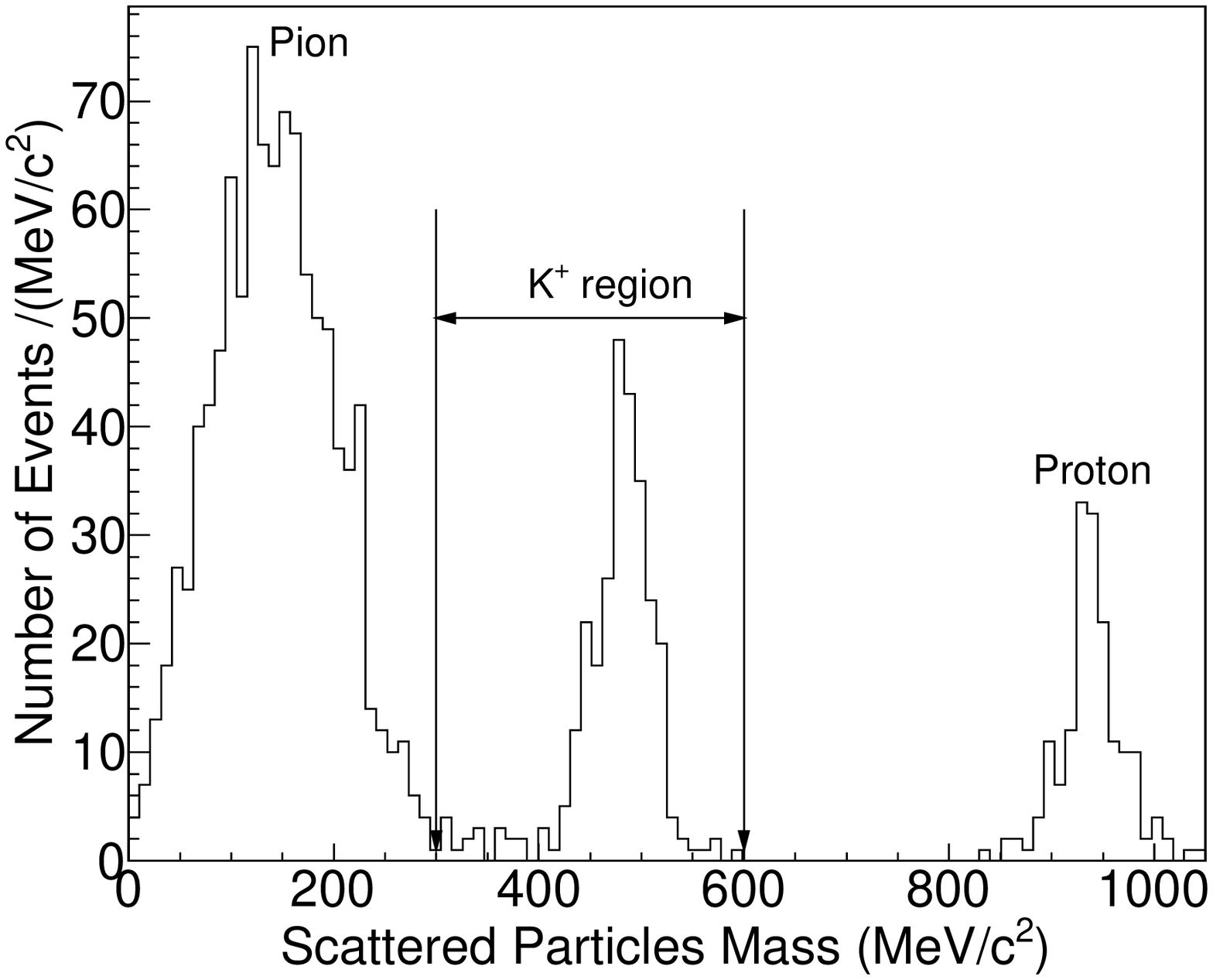}
}
\subfigure[Scattered $K^{+}$ mass spectra of true and accidental events of 
REGION-X.]{
  \label{fig:kmass_pkhigh}
  \includegraphics[height=50mm]{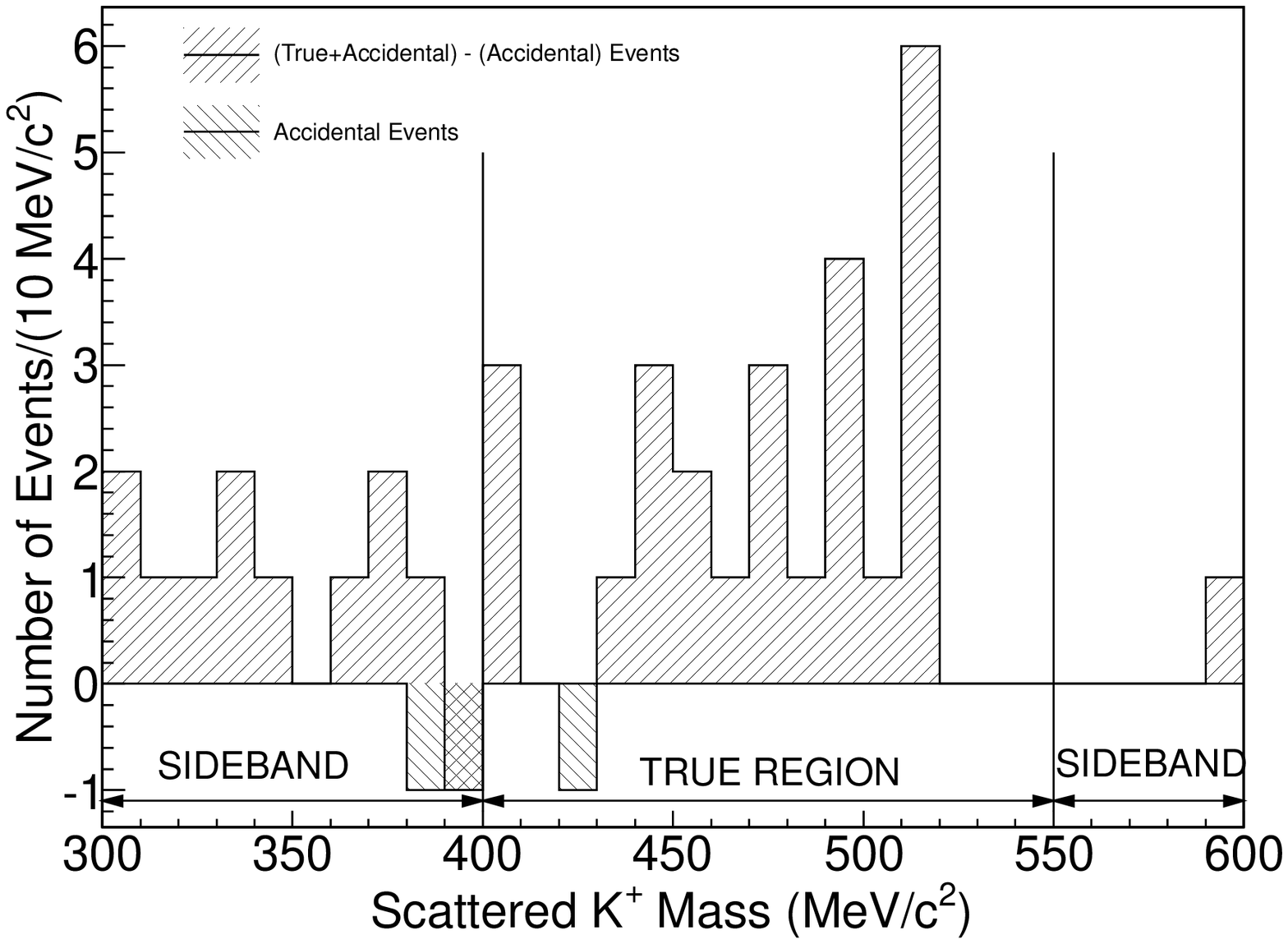}
}
 \end{center}
 \caption{
(a)Scattered particles mass spectra of the $(K^+, K^+)$ reaction on the CH$_2$ 
target gated by the $\gamma$ rays.  Contribution from the accidental 
coincidence is subtracted.  (b)The mass spectrum in $K^+$ region for 
events in REGION-X.  The true events are plotted in the positive Y-direction 
and accidental events are in the negative Y-direction. The number 
of events are used in the equation \ref{eq:signal}.
 }
\label{fig:kmass}
\end{figure}

So far, no process can explain extra events in the REGION-X in the $\gamma$ 
gated spectrum.  We further looked at the mass spectrum of scattered particles 
gated by $\gamma$ rays shown in figure \ref{fig:signal_mass}.  Since we can 
see the distinct $K^+$ peak, M$_s$=300-600 MeV/c$^2$ region was selected to 
obtain momentum spectra in figure \ref{fig:x_sim_data} and \ref{fig:cp_sim_data}.  
Mass spectrum of events in the REGION-X is also plotted in figure 
\ref{fig:kmass_pkhigh}.  The $K^+$ peak is less distinct thus we conclude 
that events in the REGION-X include backgrounds.  

In order to extract number of candidate events corresponding to the production 
of the $X$, we take the following procedure.  We define true and side band for 
both time spectrum and mass spectrum for events in the REGION-X.  T$_T$ for true 
coincidence has 5 nsec width and T$_S$ for side band has 2.5 nsec width for both 
side of true coincidence.  M$_T$ for true kaon corresponds to 400-550 MeV/c$^2$ 
and M$_S$ for side band corresponds to 300-400 MeV/c$^2$ and 550-600 MeV/c$^2$.  
Here we assume that T$_T$ and M$_T$ include background events that can be 
estimated by T$_S$ and M$_S$, respectively.  Then the number of true event 
corresponding to $X$ production ($N_X$) can be estimated by the the 
following equation, 
\begin{eqnarray}
N_{X} &=& N(T_T \cap M_T) - N(T_T \cap M_S) - N(T_S \cap M_T) + N(T_S \cap M_S) 
\nonumber \\
&=& 26 - 13 - 1 + 2 \pm \sqrt{26+13+1+2} = 14 \pm 6.48 
\label{eq:signal}
\end{eqnarray}
Here $N()$ stands for number of events for condition represented in parenthesis.  
There are two ways to show this result.  If $X$ is produced, we obserevd 
2 $\sigma$ effect.  If it is taken to be an upper limit, number of events 
corresponding to $X$ production is less than 20.5 events (1 $\sigma$).

We convert this number of events into $R_X$ which is a ratio of $X$ 
production cross section to that of $\pi^0$.  We made the following 
assumption to estimate it.  Since small momentum transfer is essential, we 
assume that all $K^+$ produced from $X^+$ is within angular acceptance of 
the spectrometer (2 $\sim$ 12 degrees) which corresponds momentum transfer 
of less than 0.1 GeV/c where $X^+$ is scattered less than 5 degrees.  On the 
other hand $\pi^0$ production follows phase 
space distribution.  Production of the $X$ gives events not only in the 
REGION-X but also below 1000 MeV/c as shown by a dashed line in figure 1. 
The number of $\pi^0$ events below 1000 MeV/c after subtraction of $X$ 
contribution and the number of $X$ events in the REGION-X were compared with 
the simulation.  We found that the $R_X = 1.0 \pm 0.5 \%$.  We can 
also give $R_X \leq 1.5\%$ as an upper limit.  The overlap of $X$ and kaon 
gives $\sim 1/100$ which indicates that the $X$ is an extended object or 
equivalently loosly bound system.  
The result suggests existence of the $X$.  Although the statistical 
significance is not enough to draw definite conclusion, it deserves 
further study.

The author (TK) is grateful for discussions with professors E. Oset, A. Hosaka 
and  H. Toki.  This work is financially supported in part by Japan Society 
for the Promotion of Science under the Japan-U.S. Cooperative Science Program.


\begin{thebibliography}{99}
  
\bibitem{TK_TS} T. Kishimoto and T. Sato, Prog. Theor. Phys.,116 (2006) 241;
hep-ex/0312003
\bibitem{nakano} T. Nakano et al., Phys. Rev. Lett. \textbf{91} (2003), 012002.
\bibitem{arndt-ps} R. A. Arndt, et al., Phys. Rev. C \textbf{68} (2003), 042201
[Errata;  \textbf{69} (2004), 019901].
\bibitem{PRC_nakano} T. Nakano et al., Phys. Rev. C \textbf{79} (2009), 025210.
\bibitem{7quark}P. Bicudo and G. M. Marques, Phys. Rev. D \textbf{69}
	(2004), 011503.
\bibitem{Estrada} F. J. Llanes-Estrada, E. Oset and V. Mateu, Phys. Rev.  
C \textbf{69} (2004), 055203.
\bibitem{Diakonov} D. Diakonov,  V. Petrov, and M. Polyakov, 
Z. Phys. A \textbf{359} (1997), 305.
\bibitem{diana}V.V. Barmin, et al., Phys. Atom. Nucl. \textbf{73} (2010) 1168
\bibitem{arndt-pin}R. A. Arndt, I. I. Strakovsky, R. L. Workman and
M. M. Pavan, Phys. Rev. C \textbf{52} (1995), 2120.
\bibitem{dover-kn} C. B. Dover and  G. E. Walker, Phys. Rep. \textbf{89} 
(1982), 1.
\bibitem{hashimoto}K. Hashimoto, Phys. Rev. C \textbf{29} (1984), 1377.
\bibitem{arndt-kn} R. A. Arndt, L. D. Roper P. H. Steinberg Phys. Rev. 
D \textbf{18} (1978), 3278.
\bibitem{estabrooks} P. Estabrooks et al., Nucl. Phys. B \textbf{133} (1978), 490.
\bibitem{e548} T. Kishimoto et al., KEK proposal E548 extension.
\bibitem{TK_PTP} T. Kishimoto et al., Prog. Theo. Phys. 118 (2007) 181    


\end{thebibliography}
\end{document}